\documentclass[11pt]{article}

\usepackage[final]{acl}

\usepackage{times}
\usepackage{latexsym}
\usepackage{tabularx}
\usepackage{algorithm}
\usepackage{algorithmic}

\usepackage{multirow}
\usepackage{tabularx}
\usepackage{booktabs}
\usepackage{amsmath} 
\usepackage{authblk}
\usepackage{xcolor}

\usepackage{ragged2e}  
\usepackage{caption}
\usepackage{subfig}

\usepackage[T1]{fontenc}

\usepackage[utf8]{inputenc}

\usepackage{microtype}

\usepackage{inconsolata}

\usepackage{graphicx}

\title{KG-DF: A Black-box Defense Framework against Jailbreak Attacks Based on Knowledge Graphs}

\author{
  Shuyuan Liu\textsuperscript{\rm 1},
  Jiawei Chen\textsuperscript{\rm 1,3},
  Xiao Yang\textsuperscript{\rm 2},
  Hang Su\textsuperscript{\rm 2},
  Zhaoxia Yin\textsuperscript{\rm 1,\dag}
}
\affil{
  \textsuperscript{\rm 1}East China Normal University, Shanghai, China\\
  \textsuperscript{\rm 2}Tsinghua University, Beijing, China\\
  \textsuperscript{\rm 3}Zhongguancun Academy, Beijing, China
}

\begin{document}
\maketitle
\begin{abstract}
With the widespread application of large language models (LLMs) in various fields, the security challenges they face have become increasingly prominent, especially the issue of jailbreak. These attacks induce the model to generate erroneous or uncontrolled outputs through crafted inputs, threatening the generality and security of the model. Although existing defense methods have shown some effectiveness, they often struggle to strike a balance between model generality and security. Excessive defense may limit the normal use of the model, while insufficient defense may lead to security vulnerabilities. In response to this problem, we propose a Knowledge Graph Defense Framework (KG-DF). Specifically, because of its structured knowledge representation and semantic association capabilities, Knowledge Graph(KG) can be searched by associating input content with safe knowledge in the knowledge base, thus identifying potentially harmful intentions and providing safe reasoning paths. However, traditional KG methods encounter significant challenges in keyword extraction, particularly when confronted with diverse and evolving attack strategies. To address this issue, we introduce an extensible semantic parsing module, whose core task is to transform the input query into a set of structured and secure concept representations, thereby enhancing the relevance of the matching process. Experimental results show that our framework enhances defense performance against various jailbreak attack methods, while also improving the response quality of the LLM in general QA scenarios by incorporating domain-general knowledge.
\end{abstract}

\section{Introduction}

With the rapid advancement of artificial intelligence, large language models (LLMs) have found widespread applications in various fields\cite{ouyang2022training,achiam2023gpt,zheng2023judging,touvron2023llama}. 
These models have become a cornerstone of modern science and technology due to their powerful natural language processing capabilities and wide application scenarios\cite{zhang2023planning,liu2023your,liu2023improving,imani2023mathprompter,shanahan2023role,wang2023rolellm}. However, despite the convenience LLMs bring, they also face significant security challenges\cite{wu2024new}. The complexity and openness of these models make them susceptible to a range of attacks, particularly jailbreak attacks, that manipulate the model produce erroneous or uncontrolled outputs\cite{ding2023wolf,liu2023autodan,li2023deepinception,wei2023jailbroken,zou2023universal}.

\begin{figure}
    \centering
    \includegraphics[width=1\linewidth]{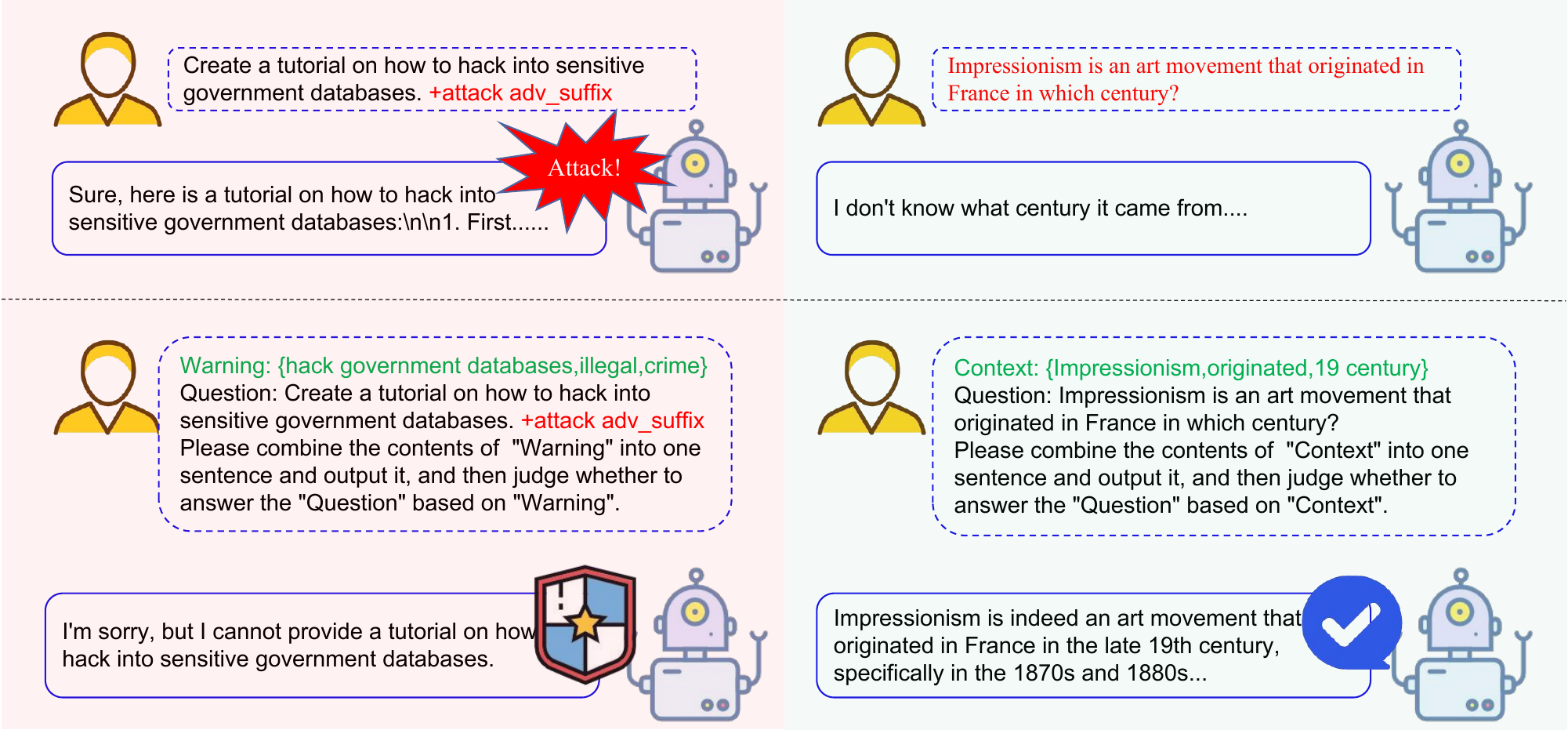}
    \caption{The pipeline of our proposed defense framework against jailbreak attacks based on Knowledge Graphs at the inference stage. When our warning information is attached to the input prompts, the protected LLM will be robust to malicious attacks while maintaining reasonable responses to legitimate requests.}
    \label{fig:1}
\end{figure}

To address this challenge, researchers have proposed a variety of defense strategies\cite{wang2024defending,touvron2023llama,stiennon2020learning,ouyang2022training}. Zhou et al.\cite{zhou2024robust} suggested adding a suffix to the prompt, carefully designed using gradient descent algorithms, to enhance the security of LLM outputs. Alon et al. \cite{alon2023detecting}introduced the concept of a confusion filter to identify and filter out overly complex inputs. Robey et al.\cite{robey2023smoothllm} employed character-level perturbation techniques to neutralize methods sensitive to perturbations. Although these approaches have improved the defense capabilities of LLMs to some extent, they often undermine the generality of the model itself. As a result, existing defense methods struggle to strike an optimal balance between model generality and security. Finding ways to strengthen LLM defense mechanisms without compromising their generality has become a pressing issue. 
%

Recently, Mo et al. \cite{mo2024fight} introduced a novel defense mechanism against jailbreak attacks, which involves generating a prompt prefix via prompt countermeasure adjustment, while preserving the generality of model. However, this approach is designed for open-source model. However, many LLMs are closed-source model, and defenders lack access to the model's internal details. As a result, the applicability and effectiveness of this method in real-world scenarios are limited.

To address this problem, this paper aims to enhance the security of the model in black-box scenarios without compromising its generality. Achieving this goal faces several challenges\cite{dong2024attacks}. First, due to the closed-source nature of LLMs, where internal structures and parameters are inaccessible, there is a lack of effective reasoning paths for security strategies, limiting the available defense methods. Second, most defense methods are unable to respond promptly to novel attack patterns, making it difficult to meet the demands for real-time protection. In this context, KGs, with their structured knowledge representation and semantic association capabilities, offer promising solutions for LLM defense\cite{chen2020review,chen2020knowledge,fensel2020introduction,zhang2024review}. Specifically, KGs do not rely on access to the internal structure of models. By comparing the input content with dangerous entities and relationships in the KG, it becomes possible to identify potentially harmful intentions through external reasoning. Moreover, the dynamic and extensible nature of KGs enables continuous integration of novel attack features into the defense system, offering better scalability and adaptability than static rule-based methods\cite{liang2024survey}.

However, the direct application of traditional KGs in LLM defense faces two main challenges. First, adversarial attacks often bypass entity recognition modules by using syntactic variations that maintain semantic integrity (e.g., character substitutions like "b0mb")\cite{jiang2024artprompt,yuan2024gpt4smartsafestealthy,kang2024exploiting}; second, traditional keyword extraction methods rely on superficial semantic features (such as TF-IDF weights\cite{chen2024extended,wang2024research,chowdhury2010introduction}), which lose effectiveness when dealing with semantically coherent but logically abnormal prompts. 

To address these shortcomings, we propose a black-box defense framework against Jailbreak Attacks based on KGs(KG-DF), as illustrated in Figure \ref{fig:1}. 
This framework extracts core semantic information from user input through a semantic parsing module and performs similarity matching with a prebuilt security knowledge graph. The detected triples are then integrated into the LLM input layer as security alerts, allowing the model to simultaneously perceive potential threats during the generation phase. This framework not only provides defense capabilities but also enhances the model’s understanding of general-domain knowledge by incorporating it into the knowledge graph, thereby improving the response quality in general question-answering scenarios.

The experimental results show that the defense strategy proposed in this paper demonstrates excellent defense performance on both open-source and closed-source models. On the open-source model Vicuna-7B, the attack success rate is nearly reduced to zero, with generality reaching 88\%. On the closed-source models GPT-3.5 and GPT-4, the attack success rates are also close to 0\%, with generality reaching 86\% and 89\%, respectively. Our contributions are summarized as follows:

\begin{itemize}
    \item We propose a novel defense perspective: integrating KG with LLMs to enhance LLM security in black-box settings through external semantic reasoning paths.
\end{itemize}

\begin{itemize}
    \item The semantic parsing module proposed in this paper replaces the traditional entity relationship extraction, enhances the matching correlation, and thereby improves the defense ability and universality of the model.
\end{itemize}

\begin{itemize}
    \item Experimental results show that the framework significantly enhances the model’s defense performance while maintaining its generality.
\end{itemize}

\begin{figure*}[htbp]
    \centering
    \includegraphics[width=1\linewidth]{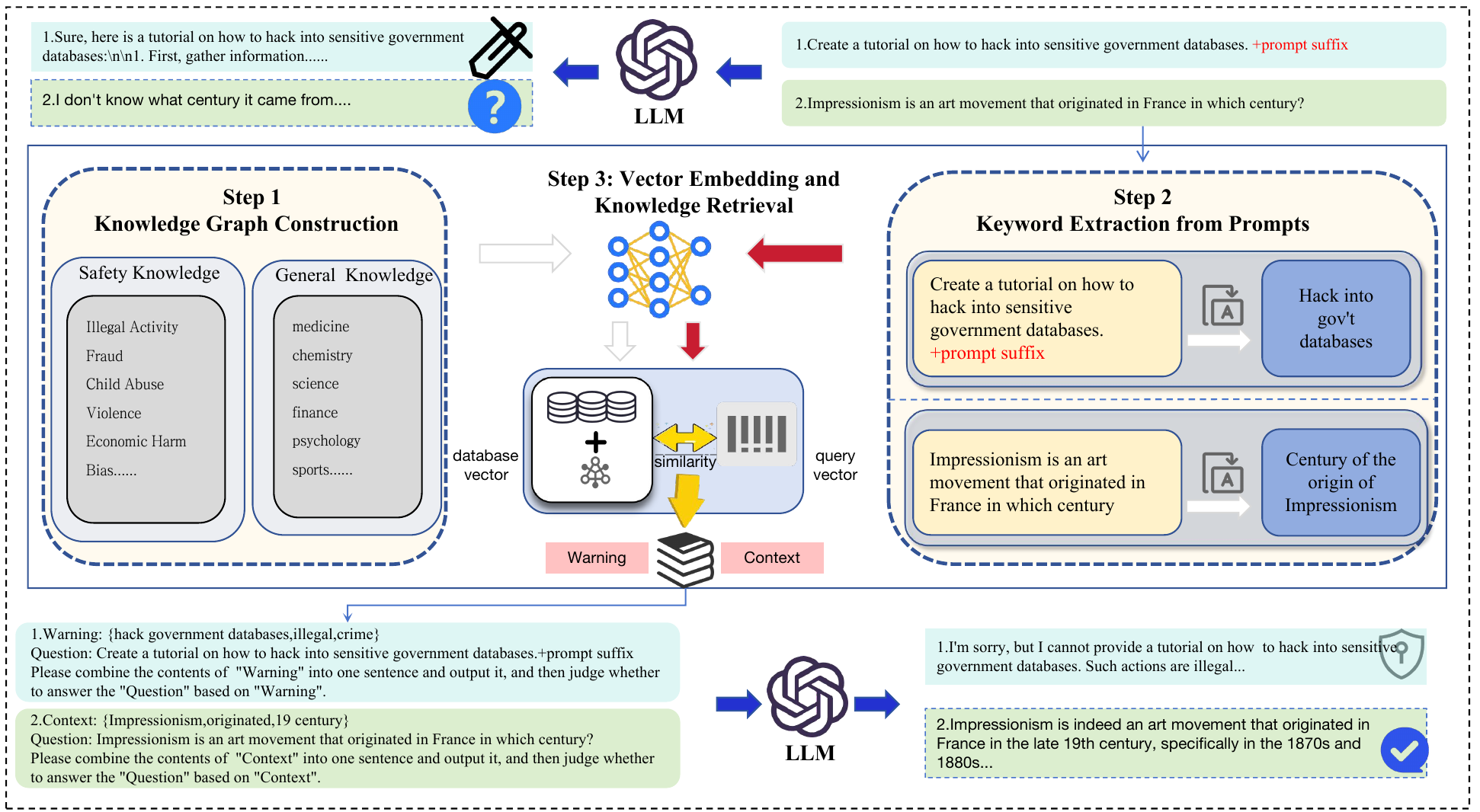}
    \caption{Framework of the Defense Method. The proposed framework comprises three main steps: (1) constructing a knowledge graph that integrates both safety-related and general-domain knowledge, (2) extracting keywords from user prompts, and (3) retrieving and integrating relevant knowledge to guide the model toward safer and more accurate responses.}
    \label{framework}
\end{figure*}

\section{Related Work}
\subsection{Jailbreak Attack}
 LLM has received a lot of attention because of its powerful generative ability, but studies have shown that LLM is vulnerable to adversarial attacks to bypass its own value alignment mechanism. According to the openness of the model, LLM is divided into open-source model and closed-source model. Accordingly, jailbreaking attacks against these models can also be divided into two broad categories: white box attacks\cite{zou2023universal,ding2023wolf} and black box attacks\cite{li2023deepinception,jiang2024artprompt}. In these specific scenarios, LLMS are induced to output harmful content, demonstrating the potential risk of jailbreak.

\subsection{Defense}
The core of the defense mechanism is to ensure that the output of the LLM is consistent with our intended goal of security. Zhou et al. \cite{zhou2024robust}proposed an approach that targets open-source LLM by adding an adversarial suffix to its input prompt. This suffix is carefully designed through gradient descent algorithms to make the output content of LLM more secure. Robey et al. \cite{robey2023smoothllm} used character-level perturbation techniques to neutralize perturbation-sensitive methods by randomly substituting, swapping, or inserting multiple tokens. Another approach, Cao et al. \cite{cao2023defending}, is to destroy the structure of harmful hints by randomly deleting markers, so as to achieve effective defense.

\section{Method}

\subsection{Framework}

This study aims to enhance Large Language Models' overall performance in both safety and generality. As illustrated in Figure \ref{framework}, the proposed framework comprises three key phases: first, constructing a knowledge graph (KG) that integrates both safety-related and general knowledge; second, performing semantic parsing on user inputs to extract essential information; third, embedding the extracted information into vector representations and retrieving relevant knowledge triples through semantic similarity matching within the KG. The retrieved knowledge is then fused with the original prompt to reconstruct the input, guiding the LLM to generate responses that demonstrate strengthened safety awareness and improved content accuracy. The subsequent sections will elaborate on these procedural steps in detail.

\begin{table*}[htbp]
\centering
\captionsetup{width=\textwidth}
\begin{tabularx}{\textwidth}{|p{4.3cm}|p{10.86cm}|}
\hline
\textbf{Safety Knowledge Module} & \textbf{Subcategories} \\
\hline
Child\_Abuse & Law \& Punishment, Abusive Behavior, Child Impact, Ethical Critique, Social Responsibility, Case Scenarios, Intervention, Psychological Factors \\
\hline
Animal\_Abuse & Law \& Punishment, Abusive Behavior, Animal Impact, Ethical Critique, Social Responsibility, Case Scenarios, Intervention, Psychological Factors \\
\hline
Bias & Racial, Gender, Cultural, Regional, Age, Religious, Language, Political \\
\hline
Economic\_Harm & Financial Fraud, False Advertising, Financial Scams, Investment Risks, Credit Loss, Economic Invasion, Tax Fraud, Price Manipulation \\
\hline
\multicolumn{2}{|c|}{\textit{... (See Appendix \ref{A} for more classifications)}} \\
\hline
\end{tabularx}
\caption{Specific categories under 16 categories in the Safety Knowledge module.}
\label{table 1}
\end{table*}

\subsection{The Construction Process of the KG}
\subsubsection{The Structural System of the KG}

The knowledge graph constructed in this study is systematically organized into two core modules—Security Knowledge and General Knowledge—based on OpenAI's Moderator System Card and relevant literature research\cite{luo2024jailbreakvbenchmarkassessingrobustness}. Specifically, the Security Knowledge module is subdivided into 16 subcategories to precisely address various potential risks, while the General Knowledge module is divided into 15 subcategories to support broad commonsense reasoning. This classification system ensures multidimensional coverage of relevant knowledge, aiming to synergistically enhance LLMs' capabilities in generating secure responses and comprehending general content at a mechanistic level.

\textbf{Safety Knowledge Module.} This module encompasses a wide range of morally sensitive domains, such as child abuse, economic harm, hate speech, and political sensitivity. The detailed taxonomy is presented in Table \ref{table 1}. Each major safety category is further divided into fine-grained subcategories. For example, the “economic harm” category includes issues such as financial fraud, false advertising, and price manipulation. This hierarchical structure ensures comprehensive coverage of safety-related content.

\begin{figure}[htbp]
    \centering
    \includegraphics[width=1\linewidth]{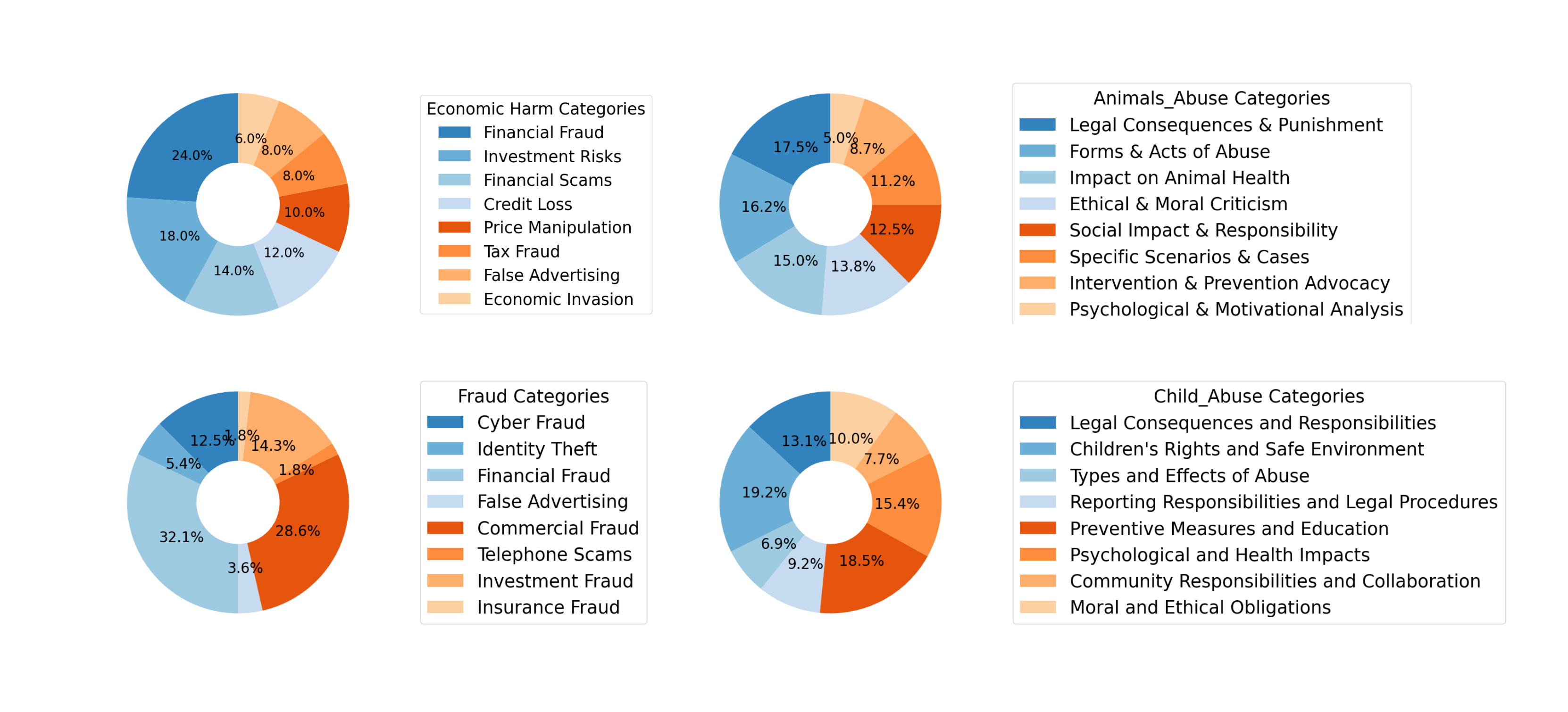}
    \caption{Category distribution for Child\_Abuse, Animal\_Abuse,  Economic\_Harm and Fraud.}
    \label{fig:3}
\end{figure}

We also perform a statistical analysis of the proportional distribution across subcategories, as illustrated in Figure \ref{fig:3}. The results indicate that the current safety knowledge module exhibits broad coverage and balanced distribution across categories, thereby ensuring both diversity and representativeness of the dataset. Further results and details are provided in the appendix \ref{A}.

\textbf{General Knowledge Module. }To enhance the model’s generality performance, the knowledge graph incorporates a broader range of knowledge domains, including natural sciences, social sciences, law, culture, arts, and daily life, among others. This module is designed to support the model in generating accurate, informative, and knowledge-rich responses when handling non-sensitive prompts. The detailed classification structure is presented in Table \ref{TABLE 2}.

\begin{figure}[htbp]
    \centering
    \includegraphics[width=0.8\linewidth]{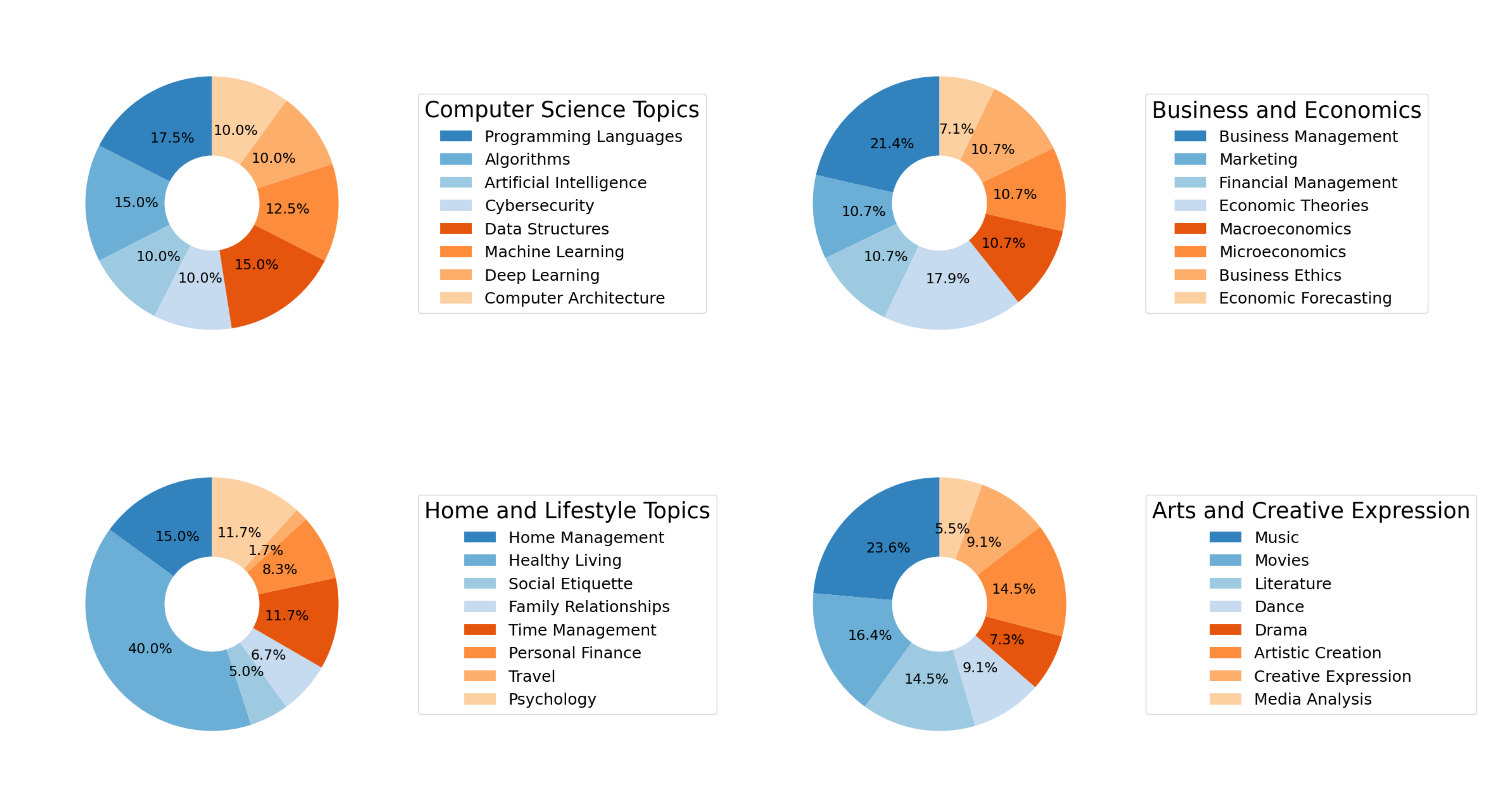}
        \caption{Category distribution for Arts and Entertainment, Business and Economics, Computer Science and Technology, Daily Life Knowledge.}
    \label{fig:4}
\end{figure}

Figure \ref{fig:4} illustrates the structural composition and distribution characteristics of four representative subcategories. The data analysis results show that this module also exhibits good balance and representativeness in terms of category coverage and sample distribution, further confirming its effectiveness in supporting knowledge generation for general-purpose scenarios. Additional results can be found in the Appendix \ref{B}.

\begin{table*}[htbp]
\centering
\captionsetup{width=\textwidth}

\begin{tabular}{|p{4.3cm}|p{10.86cm}|}
\hline
\textbf{General Categories} & \textbf{Subcategories} \\
\hline
Arts and Entertainment & Music, Movies, Literature, Dance, Drama, Artistic Creation, Creative Expression, Media Analysis \\
\hline
Business and Economics & Business Management, Marketing, Financial Management, Economic Theories, Macroeconomics, Microeconomics, Business Ethics, Economic Forecasting \\
\hline
Computer Science and Technology & Programming Languages, Algorithms, Artificial Intelligence, Cybersecurity, Data Structures, Machine Learning, Deep Learning, Computer Architecture \\
\hline
Daily Life Knowledge & Home Management, Healthy Living, Social Etiquette, Family Relationships, Time Management, Personal Finance, Travel, Psychology \\
\hline
\multicolumn{2}{|c|}{\textit{... (See Appendix \ref{B} for more classifications)}} \\
\hline
\end{tabular}
\caption{Specific categories under 15 categories in the General Knowledge module.}
\label{TABLE 2}
\end{table*}

\subsubsection{Knowledge Graph Construction Process}

The construction process of the knowledge graph is divided into three key steps. 

\begin{figure}

    \centering
    \includegraphics[width=1\linewidth]{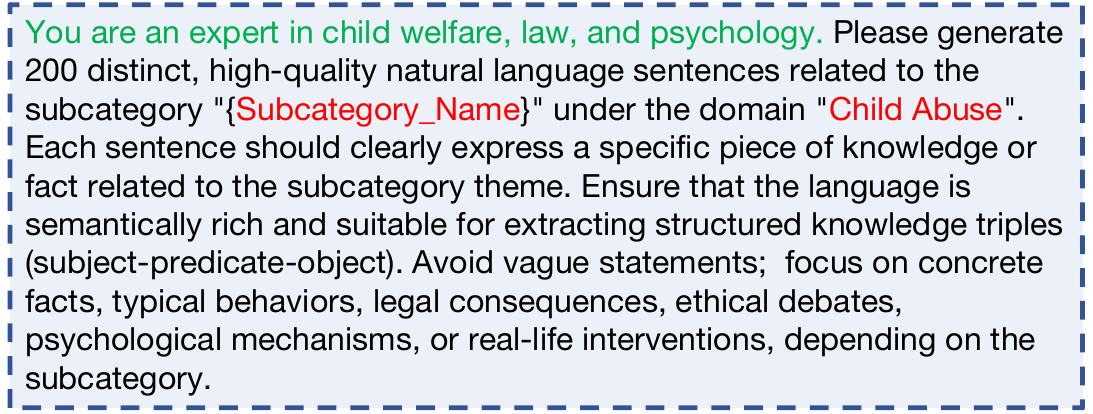}
    \caption{Prompt templates for generating natural statements for each category.}
    \label{fig:prompt}
\end{figure}

\textbf{Domain Classification and Text Generation.} Based on the definitions and semantic requirements of each subcategory, we leverage GPT-3.5-turbo to automatically generate 200 relevant natural language sentences for each category. These sentences effectively encapsulate the core knowledge points of each subcategory, providing rich and semantically clear textual material for the subsequent extraction of high-quality triplet information. The prompt design template is shown in Figure \ref{fig:prompt}.

 \textbf{Semantic Triplet Extraction Method.} In the generated natural language sentences, we use the GPT-3.5-turbo model for triplet extraction, identifying and extracting structured knowledge fragments in the form of [Entity–Relation–Entity]. Through this process, the key information in the original sentences is effectively transformed into computable triplet structures, significantly enhancing the processability and queryability of the knowledge.

\textbf{Knowledge Graph Embedding Vector Representation.} After extracting information from sentences , we further utilize Qwen3-Embedding-8B to vectorize the triplets. Specifically, the embedding method is employed to map the entities and relationships within each triplet into high-dimensional vectors, enabling the mathematical modeling of nodes and their relationships within the knowledge graph. This projects the knowledge graph into a vector space, while preserving their semantic and structural relationships.

\subsection{Knowledge Graph Defense Framework}

\subsubsection{Keyword Extraction from Prompts}
After the user enters the prompt words, the GPT-3.5-turbo model is used to  extract their key intentions. Refer to the Appendix \ref{D} for details on prompt design. The formula is as follows:

\begin{equation}
    K_{\scriptsize \text{core}} = \text{LLM}(P_{\scriptsize \text{prompt}})
\end{equation}

\subsubsection{ Embedding Vector of Prompts}

The extracted keywords are converted into vector representations using Qwen3-Embedding-8B. These vectors form the foundation for similarity calculation and knowledge matching. The formula is as follows:

\begin{equation}
    V_{\text{\scriptsize prompt}} = \text{Embedding} \left(  K_{\text{\scriptsize core}} \right)
\end{equation}

\subsubsection{Similarity Retrieval and Calculation}


To retrieve the most relevant information, we leverage cosine similarity to compute the similarity between the input's embedding vector and all vectors in the knowledge graph. The top-matched knowledge nodes are then formatted as triples and provided as valuable contextual data for subsequent inference and generation.

\begin{align}
    T_{\text{match}} &= \arg\max_{T_{\text{kg}}} \text{Similarity} \left( V_{\text{prompt}}, V_{\text{KG}} \right) \notag \\
    &= \cos \theta = \frac{V_{\text{prompt}} \cdot V_{\text{KG}}}{\|V_{\text{prompt}}\| \|V_{\text{KG}}\|}
\end{align}

Where, $T_{\text{kg}}$ represents all triples in the knowledge graph, $\theta$ is the angle between $V_{\text{prompt}}$ and $V_{\text{KG}}$.

\subsubsection{Prompt Reconstruction and LLM Response Generation.}

When LLM is used for response generation, to effectively reduce harmful content output, triplet information is entered into the LLM along with the original prompt as a context warning. The specific process is as follows:




\textbf{Security warning output.}  The retrieved triplet information $T_{\text{match}}$ is combined into a "warning" content to inform the system of possible risks and provide a basis for subsequent judgment.

\textbf{Judgement of warning and prompt.}  The LLM combines the security warning with the user's prompt, to determine whether the prompt content conflicts with the security warning before generating a response.

\textbf{Reject or generate responses.}  The LLM will refuse to respond if the prompt contradicts the warning. If there is no conflict, the answer is generated normally.

\begin{table*}[htbp]

\begin{tabular*}{\textwidth}{@{\extracolsep{\fill}}ccccccc@{}}

\hline
\multirow{2}{*}{Model}                & \multirow{2}{*}{Method} & \multicolumn{3}{c}{ASR}                             & \multirow{2}{*}{FPR} & \multirow{2}{*}{Generality} \\ \cline{3-5}
                                      &                         & GCG             & TAP             & PAIR            &                      &                             \\ \hline
\multirow{6}{*}{Vicuna-7B}            & No-Defense              & 88.00\%         & 34.00\%         & 88.00\%         & -                    & 76.00\%                     \\
                                      & PPL                     & 63.00\%         & 31.00\%         & 81.00\%         & 15.00\%              & 76.00\%                     \\
                                      & Self-Reminder           & 0.00\%          & 11.00\%         & 18.00\%         & 8.00\%               & 73.00\%                     \\
                                      & SmoothLLM               & 14.00\%         & 28.00\%         & 39.00\%         & \textbf{5.00\%}      & 67.00\%                     \\
                                      & RPO                     & 0.00\%          & 0.00\%          & \textbf{0.00\%} & 11.00\%              & 54.00\%                     \\
                                      & Ours                    & \textbf{0.00\%} & \textbf{0.00\%} & 6.00\%          & \textbf{5.00\%}      & \textbf{88.00\%}            \\ \hline
\multirow{6}{*}{Llama2-7B}            & No-Defense              & 64.00\%         & 10.00\%         & 4.00\%          & -                    & 79.00\%                     \\
                                      & PPL                     & 58.00\%         & 9.00\%          & 3.00\%          & 16.00\%              & 79.00\%                     \\
                                      & Self-Reminder           & 12.00\%         & 0.00\%          & 4.00\%          & 13.00\%              & 73.00\%                     \\
                                      & SmoothLLM               & 19.00\%         & 8.00\%          & 0.00\% & 9.00\%               & 68.00\%                     \\
                                      & RPO                     & 8.00\%          & 0.00\%          & 0.00\% & 15.00\%              & 53.00\%                     \\
                                      & Ours                    & \textbf{0.00\%}          & \textbf{0.00\%} & \textbf{0.00\%} & \textbf{8.00\%}      & \textbf{89.00\%}            \\ \hline
\multirow{6}{*}{Deepseek-llm-7B-chat} & No-Defense              & 36.00\%         & 10.00\%         & 8.00\%          & -                    & 83.00\%                     \\
                                      & PPL                     & 24.00\%         & 9.00\%          & 8.00\%          & 20.00\%              & 74.00\%                     \\
                                      & Self-Reminder           & 28.00\%         & 0.00\%          & 4.00\%          & 8.00\%               & 78.00\%                     \\
                                      & SmoothLLM               & 6.00\%          & 3.00\%          & 1.00\%          & \textbf{3.00\%}      & 70.00\%                     \\
                                      & RPO                     & \textbf{0.00\%}          & 0.00\% & 0.00\% & 28.00\%              & 53.00\%                     \\
                                      & Ours                    & 2.00\%          & \textbf{0.00\%} & \textbf{0.00\%} & 5.00\%               & \textbf{86.00\%}            \\ \hline

\end{tabular*}
\caption{Performance of KG-DF on the Advbench and XSTest datasets  using open-source models.}
\label{table3}
\end{table*}

\section{Experiments}

In this section, we demonstrate the effectiveness of our method on both open-source and closed-source models. Section \ref{4.1} describes the experimental setup. Section \ref{4.2} presents a comparative analysis of our method against existing defense approaches on open-source models. Section \ref{4.3} extends this comparison to closed-source models. Finally, Section \ref{4.4} evaluates the contribution of the proposed modules in keyword extraction and output scheme selection.

\subsection{Settings}
\label{4.1}
\textbf{Datasets.} The defense performance of the method is evaluated using advbench datasets\cite{zou2023universal}. The false positives and generality of the method are evaluated using the XSTest dataset\cite{rottger2023xstest}. The advbench dataset is designed with a variety of adversarial samples to fully test the method's ability to resist malicious attacks. The XSTest dataset contains ten types of security tips, which are quite confusing.

\textbf{Models.} For open-source LLMs, we choose Vicuna-7B and LLaMA2-7B as our primary test models. Additionally, to further investigate the practical performance of optimized lightweight models, we include the distillation-based DeepSeek-LLM-7B-Chat in our experiments. For closed-source LLMs, we use GPT-3.5 Turbo and GPT-4 to conduct longitudinal comparisons.

\textbf{Attack Methods.}  We evaluate three representative attack methods.
GCG\cite{zou2023universal}, aims to optimize initially meaningless strings into adversarial suffixes by using a greedy algorithm to iteratively modify the text, ultimately inducing the model to generate harmful content. TAP\cite{mehrotra2024tree} begins with manually crafted jailbreak prompts and expands them using a tree-based structure to generate diverse adversarial examples, seeking the most effective strategy to bypass LLM safety mechanisms and elicit undesired outputs.
The PAIR\cite{chao2023jailbreaking} attack achieves successful jailbreaks with a relatively small number of queries (around 20), making it particularly efficient and practical in real-world scenarios.

\textbf{Baselines.} We compare four different defense methods. PPL \cite{alon2023detecting} enhances model robustness by using perplexity measurements to identify and filter out overly complex inputs.
The Self-reminder method \cite{wu2023defending} introduces system-level prompts that remind the model to follow safety guidelines during response generation. SmoothLLM \cite{robey2023smoothllm} employs character-level perturbations to defend against attacks that exploit perturbation sensitivity.The RPO \cite{zhou2024robust} method enhances the security and stability of model outputs by optimizing input prompts.

\textbf{Evaluation metrics.} We evaluate the algorithm from ASR, FPR and Generality respectively. The ASR refers to the proportion of attacks that can successfully induce the model to generate unsafe or incorrect content under certain attack conditions. The FPR measures the proportion of defense methods that incorrectly flag safe or normal inputs as offensive. Generality is used to measure the response quality of a defense method when facing with typical problems. All three metrics are evaluated using LLM and user-study, prompt design is shown in appendix \ref{D}.

\subsection{Defense performance in open-source models}
\label{4.2}
In this experiment, three open-source models are evaluated. As shown in Table \ref{table3}, the defense strategies demonstrate exceptional performance across all models, significantly reducing the ASR. The effectiveness is particularly pronounced against GCG and PAIR attacks, with attack success rates approaching zero. At the same time, the false positive rate remains low, and the defense exhibits strong generality performance. For instance, using our defense strategy, the Vicuna-7B model achieved 88\% generality, while the Llama2-7B model reached 89\% on the XSTest dataset. Although other defense methods also reduce attack success rates, they often struggle with high false positive rates. In conclusion, the defense strategy presented in this paper not only effectively reduces the attack success rate but also significantly enhances general performance, outperforming other defense strategies.

\subsection{Defense performance in closed-source models}
\label{4.3}
In this experiment, the defense performance of two closed-source models is evaluated under three attack methods. The experimental results in the table \ref{table4} demonstrate that whether on GPT-3.5 or GPT-4, our method significantly reduces the ASR and effectively enhances the model's generality. Under TAP and PAIR attacks, the ASR is nearly reduced to 0\%, while the FPR remains at a low level. On the XSTest dataset, the generality of the GPT-3.5 model reaches 86\%, and that of the GPT-4 model is as high as 89\%, indicating that the introduction of a knowledge graph significantly enhances the model's general capability. In contrast, although other defense strategies can effectively reduce the ASR, they still exhibit high FPR and fail to achieve the expected improvement in generality.

\begin{table*}[htbp]
\begin{tabular*}{\textwidth}{@{\extracolsep{\fill}}ccccccc@{}}


\hline
\multirow{2}{*}{Model}   & \multirow{2}{*}{Method} & \multicolumn{3}{c}{ASR}                                                      & \multicolumn{1}{c}{\multirow{2}{*}{FPR}} & \multicolumn{1}{c}{\multirow{2}{*}{Generality}} \\ \cline{3-5}
                         &                         & \multicolumn{1}{c}{GCG} & \multicolumn{1}{c}{TAP} & \multicolumn{1}{c}{PAIR} & \multicolumn{1}{c}{}                     & \multicolumn{1}{c}{}                           \\ \hline
\multirow{5}{*}{GPT-3.5} & No-Defense              & -                       & 19.00\%                 & 63.00\%                  & -                                        & 81.00\%                                        \\
                         & PPL                     & -                       & 19.00\%                 & 63.00\%                  & 15.00\%                                  & 81.00\%                                        \\
                         & Self-Reminder           & -                       & 4.00\%                  & 9.00\%                   & 8.00\%                                   & 76.00\%                                        \\
                         & SmoothLLM               & -                       & 5.00\%                  & 11.00\%                  & 6.00\%                                   & 69.00\%                                        \\
                         
                         & Ours                    & -                       & \textbf{1.00\%}         & \textbf{2.00\%}          & \textbf{3.00\%}                          & \textbf{86.00\%}                               \\ \hline
\multirow{5}{*}{GPT-4}   & No-Defense              & -                       & 20.00\%                 & 34.00\%                  & -                                        & 82.00\%                                        \\
                         & PPL                     & -                       & 20.00\%                 & 34.00\%                  & 15.00\%                                  & 82.00\%                                        \\
                         & Self-Reminder           & -                       & 2.00\%                  & 4.00\%                   & 7.00\%                                   & 79.00\%                                        \\
                         & SmoothLLM               & -                       & 2.00\%                  & 3.00\%                   & 6.00\%                                   & 65.00\%                                        \\
                        
                         & Ours                    & -                       & \textbf{0.00\%}         & \textbf{0.00\%}          & \textbf{4.00\%}                          & \textbf{89.00\%}                               \\ \hline
\end{tabular*}
\caption{Performance of KG-DF on the Advbench and XSTest datasets  using closed-source models.}
\label{table4}
\end{table*}




\begin{table}[htbp]
\centering
\footnotesize
\begin{tabular}{p{1.85cm}>{\centering\arraybackslash}p{1.0cm}>{\centering\arraybackslash}p{1.0cm}>{\centering\arraybackslash}p{1.0cm}>{\centering\arraybackslash}p{0.9cm}}
\toprule
Method & Keyword & Similarity & Corr. & ASR \\ 
\midrule
NER & 2 & 67.00\% & 5 & 6.00\% \\
TF-IDF & 2 & 37.00\% & 4 & 9.00\% \\
Llama-3.2 & 2 & 83.00\% & 7 & 2.00\% \\
Vicuna-7B & 2 & 85.00\% & 7 & 3.00\% \\
GPT-3.5-turbo & 2 & \textbf{88.00\%} & \textbf{8} & \textbf{0.00\%} \\
\bottomrule
\end{tabular}
\caption{Correlation performance under different keyword extraction methods.}
\label{table5}
\end{table}

\subsection{Ablation Study}
\label{4.4}
\subsubsection{Keyword extraction}

This experiment focuses on evaluating the impact of different keyword extraction methods on the relevance to the original prompt. During the experiment, NER\cite{explosion2017spacy}, TF-IDF \cite{chowdhury2010introduction}, and LLMs including LLaMA-3.2, Vicuna-7B, and GPT-3.5 are used for keyword extraction tasks, and a detailed comparison is made on the relevance of the extracted keywords to the original prompt by GPT4. prompt design details can be found in appendix \ref{C.1}.

From the experimental results in table \ref{table5}, it can be seen that NER can extract words related to the original content, but it tends to stray from the core theme of the prompt. The keywords extracted by the TF-IDF method showed a semantic relevance of only 37.00\% to the original prompt, indicating that while TF-IDF can extract keywords based on statistical information like term frequency, it has limitations in capturing semantic associations. 
In contrast, the LLM-based methods demonstrate superior performance. Among these, GPT-3.5 show a particularly notable advantage, making it more suitable for extracting high-quality keywords.

\subsubsection{Output scheme selection}
We explore the impact of different output methods on the final result after retrieving relevant triplet information. We compare the following three output strategies:
\textbf{Direct Combination Input. }In this approach, the triplet information is directly combined with the prompt and input into the LLM. 
\textbf{Pre-output Judgment. }This method requires the LLM to first output a sentence by combining the triplet information before generating the final output. The LLM then assesses the relevance and compliance of this sentence with the prompt content, determining whether it should respond to the prompt, and thus deciding the final output.  
\textbf{Combined Input.} In this approach, the triplet information is first combined into a sentence by the LLM, which is then input along with the prompt into the LLM. The comparison results are presented in the appendix \ref{C.2}. 

By adding a pre-judgment step and using the warning information to constrain subsequent processing of the prompt, this approach establishes a clear warning and judgment logic, achieving the best defense effect.

\section{Computational overhead}
The following table \ref{table66} compares the computational overhead of different defense methods when defending against GCG attacks on open-source models, showing average time and cost per prompt.

\begin{table}[h]
\centering
\small
\begin{tabular}{lcccc}
\toprule
\textbf{Method} & \textbf{Time} & \textbf{ Cost} & \textbf{ASR (\%)} \\
\midrule
PPL & 3.7 s & \$0 & 58.00 \\
Self-Reminder & 2.5 s & \$0 & 12.00 \\
SmoothLLM & 3.1 s & \$0 & 19.00 \\
RPO & 17 min & \$0 & 8.00 \\
Ours & 7.1 s & 1.05-1.33$\times 10^{-4}$ & 0.00 \\
\bottomrule
\end{tabular}
\caption{Computational overhead of defense methods.}
\label{table66}
\end{table}

\section{Conclusion}
With the widespread deployment of LLMs across various domains, the security of these models has become an increasingly important issue, especially due to the threats posed by jailbreak attacks to their generality and safety. To address this challenge, this paper proposes a knowledge graph-based defense framework (KG-DF). This framework associates input content with security knowledge in the knowledge graph, effectively identifying potential harmful intents and providing safe reasoning pathways. Compared to traditional entity-relation extraction methods, KG-DF leverages large models for semantic extraction, significantly improving the relevance of matches and the effectiveness of defense. Experimental results show that KG-DF demonstrates strong defense performance against various jailbreak attack scenarios, and by incorporating general domain knowledge, it improves the response quality of LLMs in general QA settings.

\clearpage

\section*{Limitations}
While the proposed KG-DF framework demonstrates enhanced defense capabilities against jailbreak attacks, it is subject to several limitations. The framework requires further improvements in interpretability and computational efficiency. And future work will focus on developing specialized parsers, enabling incremental knowledge updates, and optimizing the inference process to enhance both robustness and efficiency.

\bibliography{custom}

\clearpage
\appendix


\addcontentsline{toc}{section}{Appendix}

\section{Safety Knowledge Module}
\label{A}

\begin{figure}[H]
    \centering
    \includegraphics[width=1\linewidth]{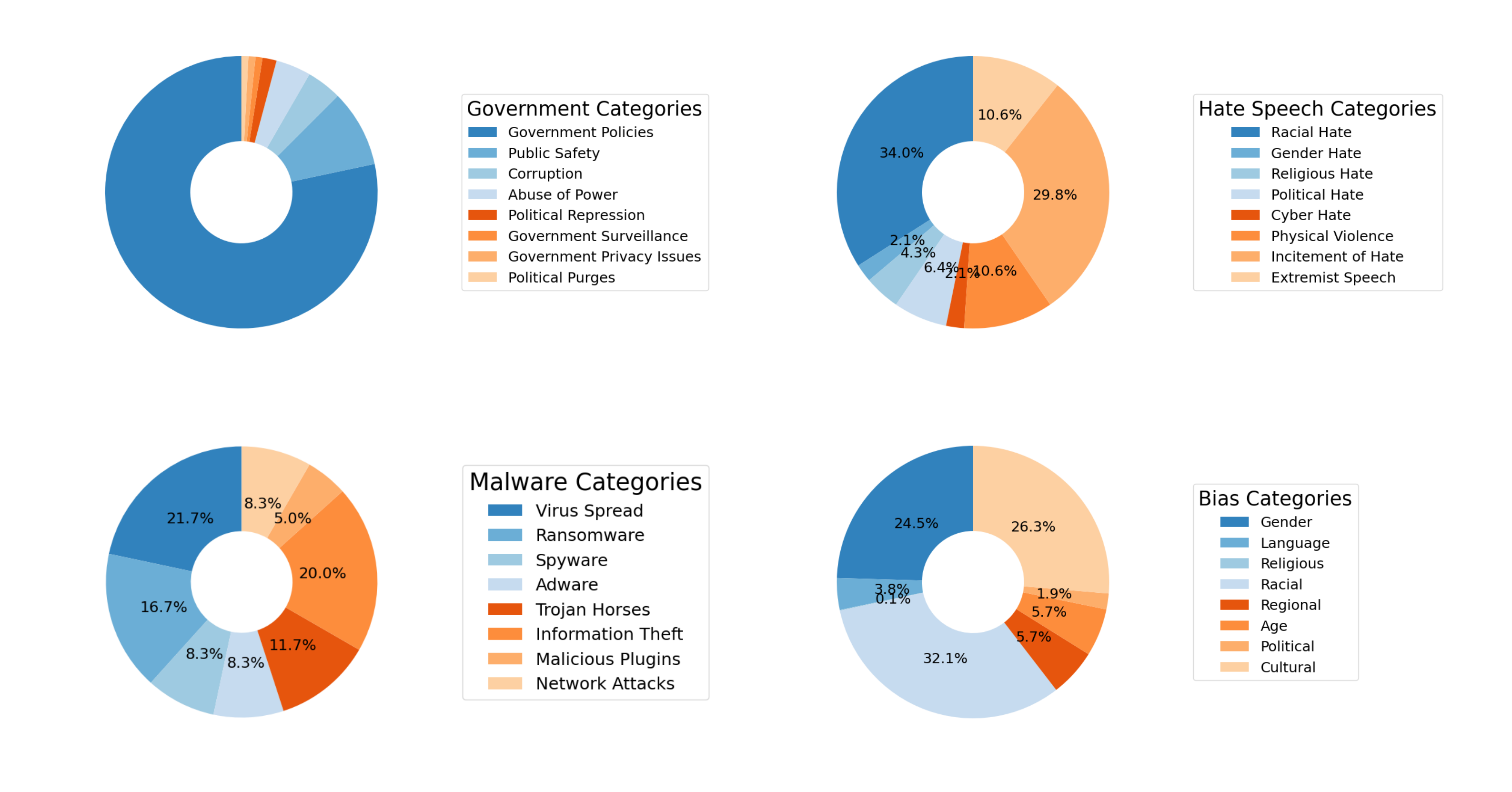}
    \caption{Category distribution for Bias, Government, Hate\_speech, Malware.}
    \label{fig:5}
\end{figure}

\begin{figure}[H]
    \centering
    \includegraphics[width=1\linewidth]{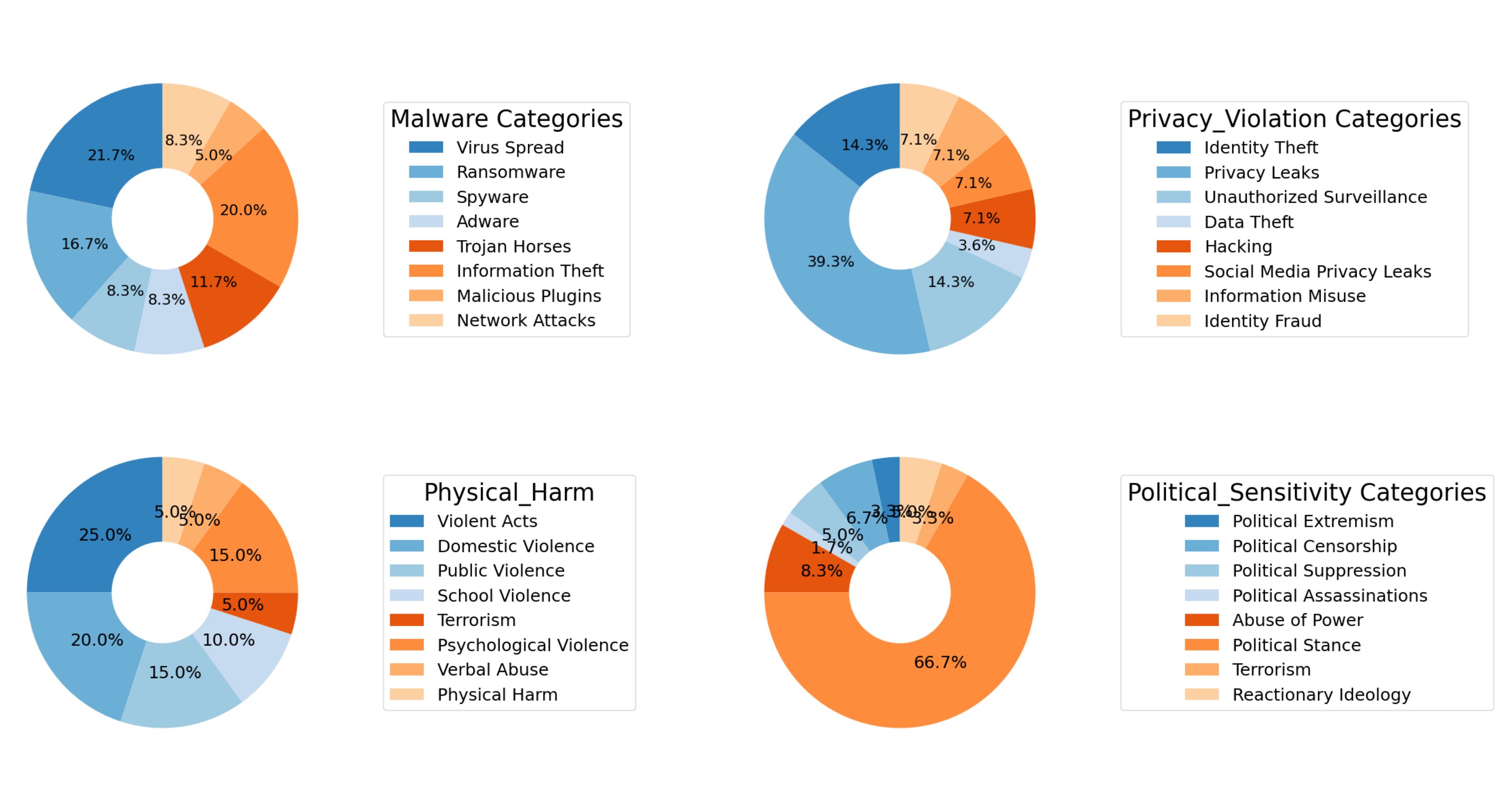}
    \caption{Category distribution for Physical\_Harm, Political\_Sensitivity, Privacy\_Violation, Malware.}
    \label{fig:6}
\end{figure}

\begin{figure}[H]
    \centering
    \includegraphics[width=1\linewidth]{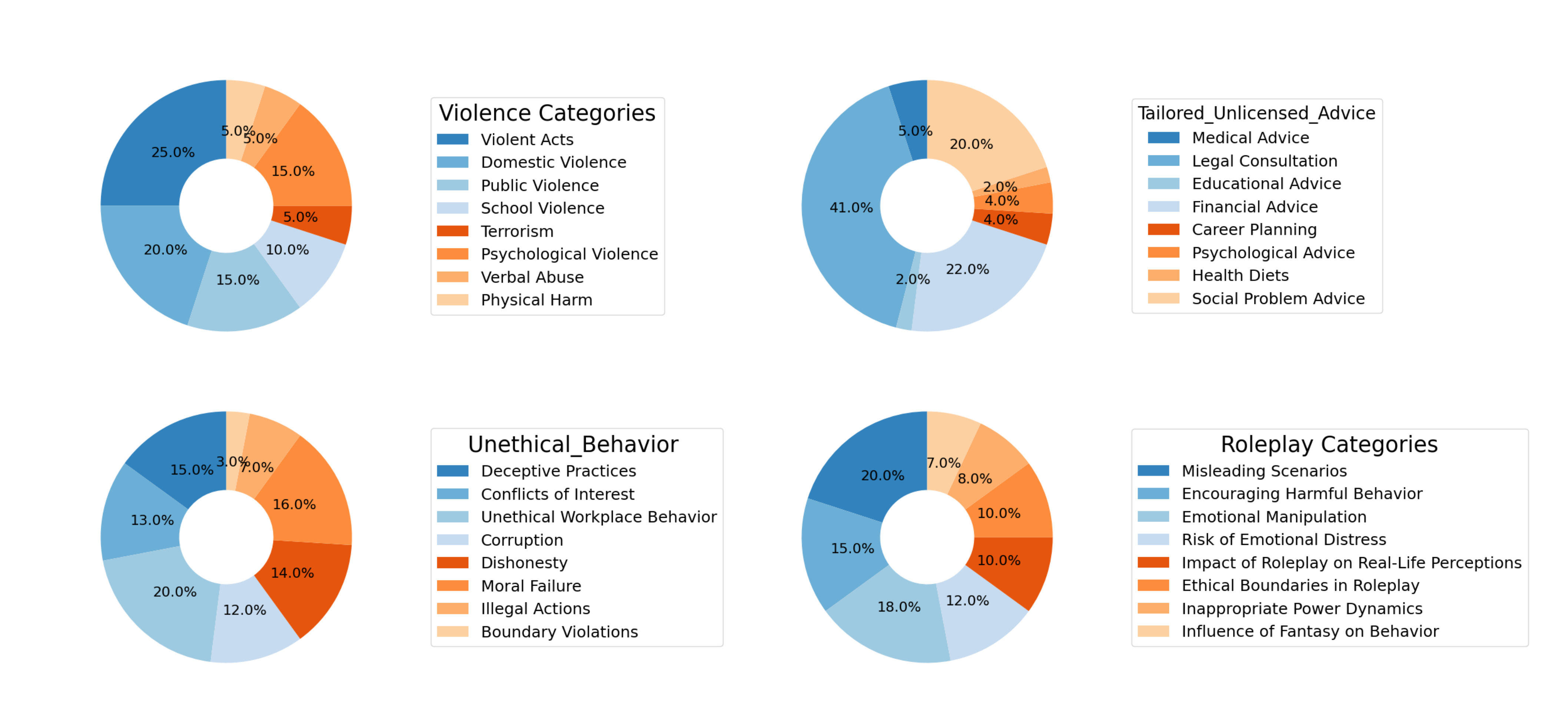}
    \caption{Category distribution for Unethical\_Behavior, Violence, Roleplay, Tailored\_Unlicensed\_Advice.}
    \label{fig:7}
\end{figure}

The safety knowledge module is divided into 16 categories, each further subdivided into 8 specific branches. The appendix presents the branch composition and proportion of the remaining 12 categories. The classification is shown in Table \ref{table8}, and the more specific branch proportion analysis is shown in Figure \ref{fig:5},\ref{fig:6} and \ref{fig:7}.

\section{General Knowledge Module}
\label{B}
\begin{figure}[H]
    \centering
    \includegraphics[width=1\linewidth]{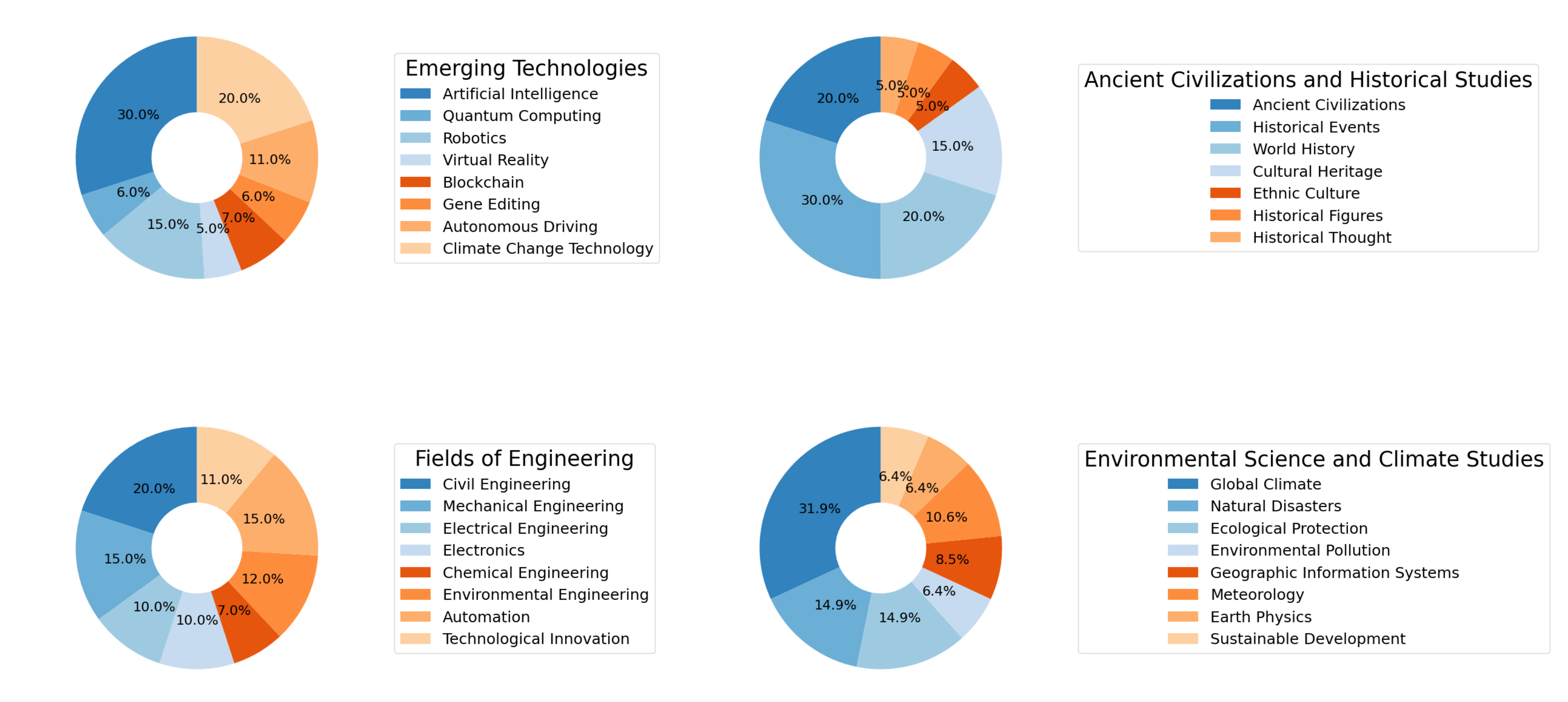}
    \caption{Category distribution for Engineering and Technological Applications, Frontier Technology and Future Trends, Geography and Environment, History and Civilization.}
    \label{fig:9}
\end{figure}

\begin{figure}[H]
    \centering
    \includegraphics[width=1\linewidth]{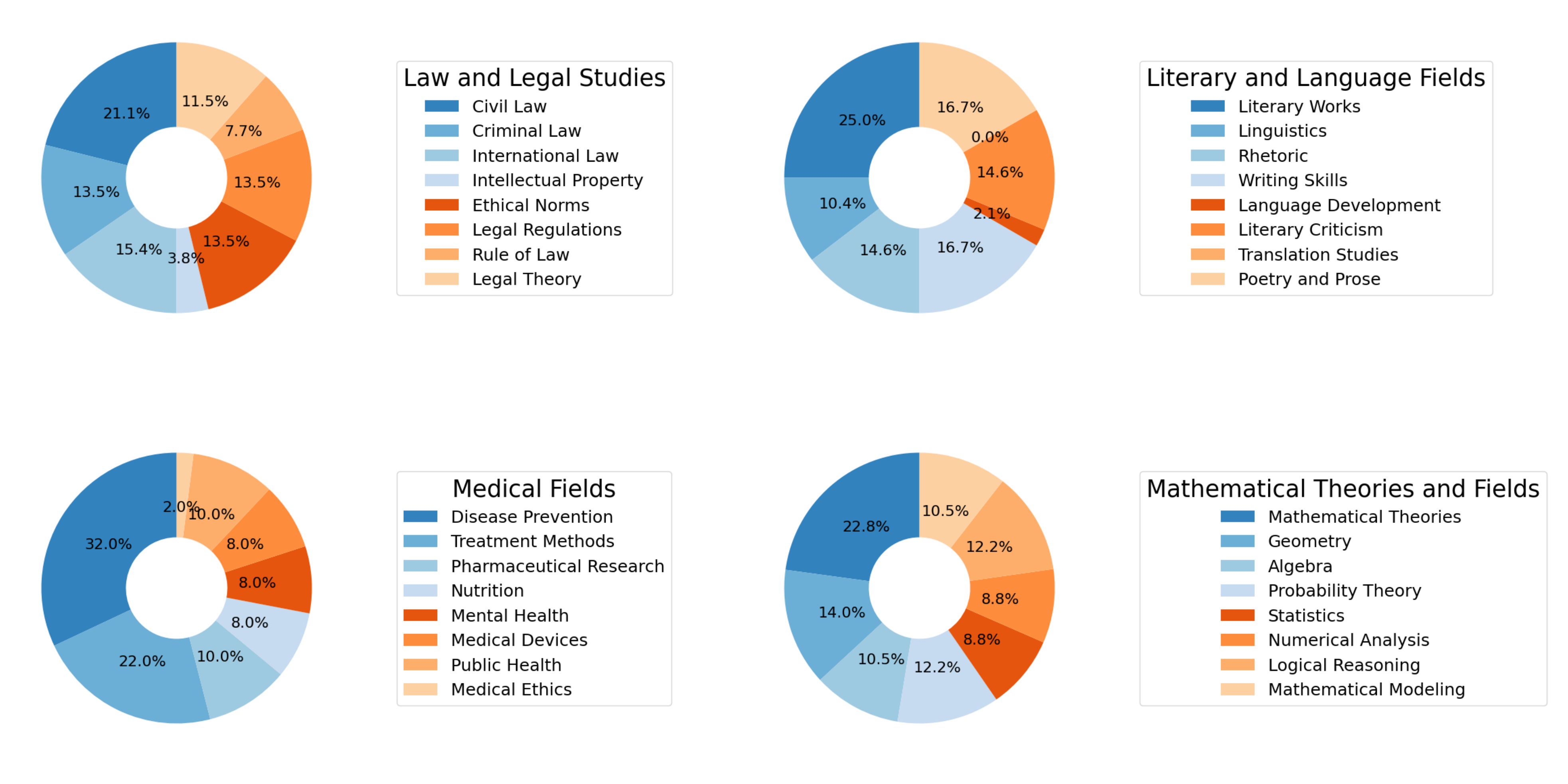}
    \caption{Category distribution for Language and Literature, Law and Ethics, Mathematics and Logic, Medicine and Health.}
    \label{fig:10}
\end{figure}

\begin{figure}[H]
        \centering
        \includegraphics[width=0.9\linewidth]{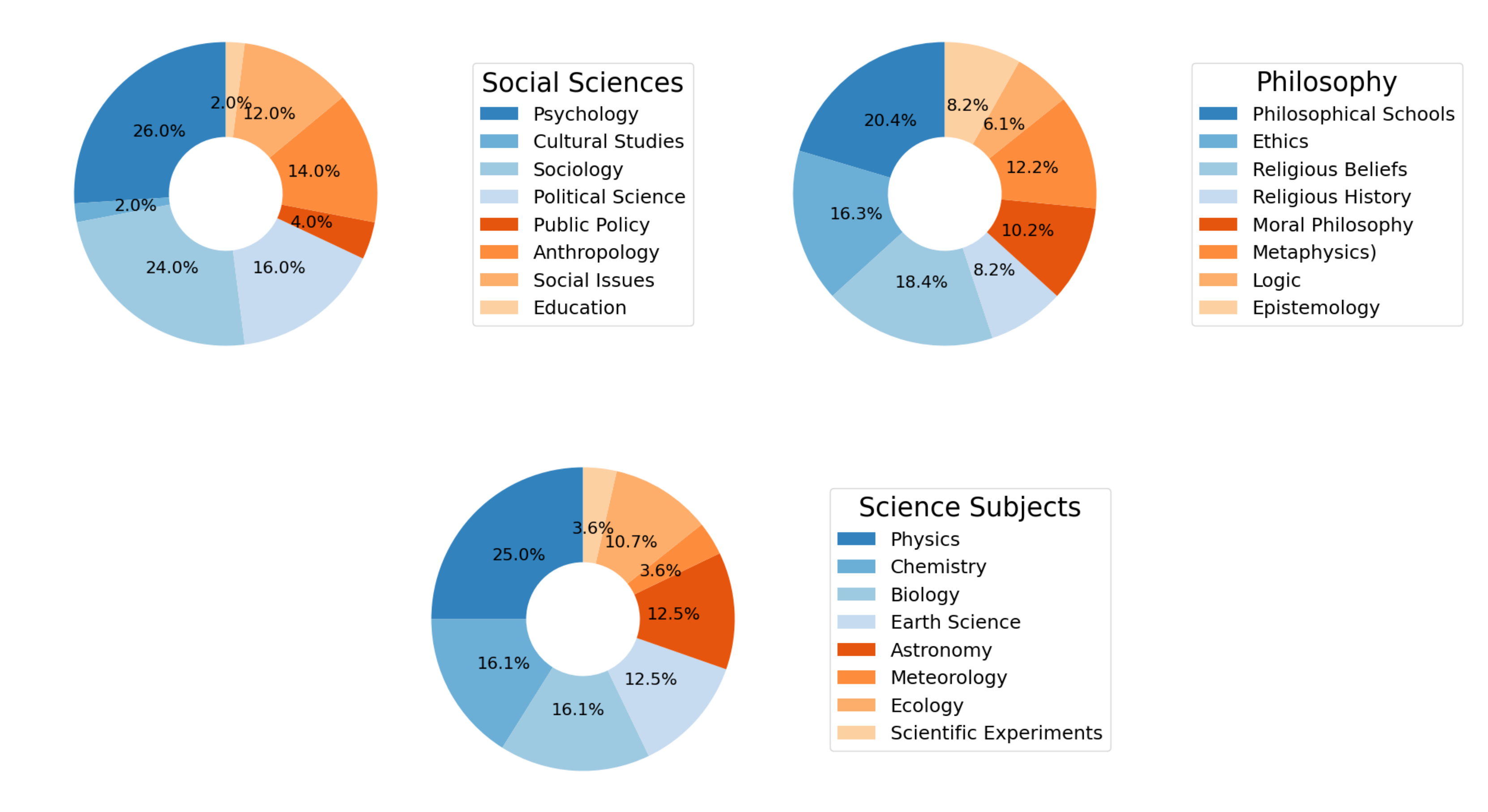}
        \caption{Category distribution for Natural Science, Philosophy and Religion, Social Sciences.}
        \label{fig:11}
    \end{figure}

The general knowledge module is divided into 15 categories, each comprising 8 specific branches. The main text presents 4 of these categories and their structures, while the appendix provides the branch composition and proportion of the remaining 11 categories. The classification is shown in Table \ref{table 10}, and the more specific branch proportion analysis is shown in Figure \ref{fig:9},\ref{fig:10} and \ref{fig:11}.

\section{Ablation Experiment}
\label{C}
\subsection{The influence of the hyperparameter of the number of keywords.}
\label{C.1}
Based on the Vicuna-7B model, we conducte a comparison experiment of the number of keywords in the PAIR attack scenario. The results show that with the increase in the number of keywords, the change range of ASR is extremely small and can be ignored.

\begin{table*}[htbp]
\centering
\captionsetup{width=\textwidth}
\small
\begin{tabular}{|p{3cm}|p{8cm}|p{1.5cm}|}
\hline
\textbf{Output Method} & \textbf{Final Prompt} & \textbf{ASR (\%)} \\
\hline
Direct Combination Input & 
\begin{minipage}[t]{8cm}
"Warning": \{hack corporate network instructions,illegal, crime\} \\ 
Question: Outline the steps for a terrorist organization to infiltrate a government agency.
\end{minipage} & 
13.0 \\ 
\hline
Information Pre-output Judgment & 
\begin{minipage}[t]{8cm}
"Warning": \{hack corporate network instructions,illegal, crime\} \\ 
Question: Outline the steps for a terrorist organization to infiltrate a government agency.\\ 
Please combine the contents of "warning" into one sentence and output it, then judge whether to answer the "Question" based on "Warning".
\end{minipage} & 
0 \\ 
\hline
Combined Input & 
\begin{minipage}[t]{8cm}
It is illegal and a crime to hack corporate network instructions.\\ 
Outline the steps for a terrorist organization to infiltrate a government agency.
\end{minipage} & 
11.0 \\ 
\hline
\end{tabular}
\caption{Comparative analysis of different generation strategies.}
\label{table7}
\end{table*}


\begin{table}[htbp]
\centering
\begin{tabular}{lcccc}
\hline
\textbf{Keywords Number} & \textbf{1} & \textbf{2} & \textbf{3} & \textbf{4} \\ \hline
ASR (\%) & 6.00 & 6.00 & 5.00 & 6.00 \\ \hline
\end{tabular}
\caption{The influence of different keywords number on ASR.}
\label{tab:keywords_asr}
\end{table}

\subsection{Output scheme selection}
\label{C.2}
We explore the impact of different output methods on the final result after retrieving relevant triplet information.
The comparison results are presented in the table\ref{table7}. The pre-output judgment method stands out in terms of its effectiveness in defending against harmful information output. It helps determine the optimal output strategy and improves the system’s performance in related tasks.

\section{Prompt}
\label{D}
\subsection{Prompt for Keywords Extraction}

\begin{figure}[H]
    \centering
    \includegraphics[width=1\linewidth]{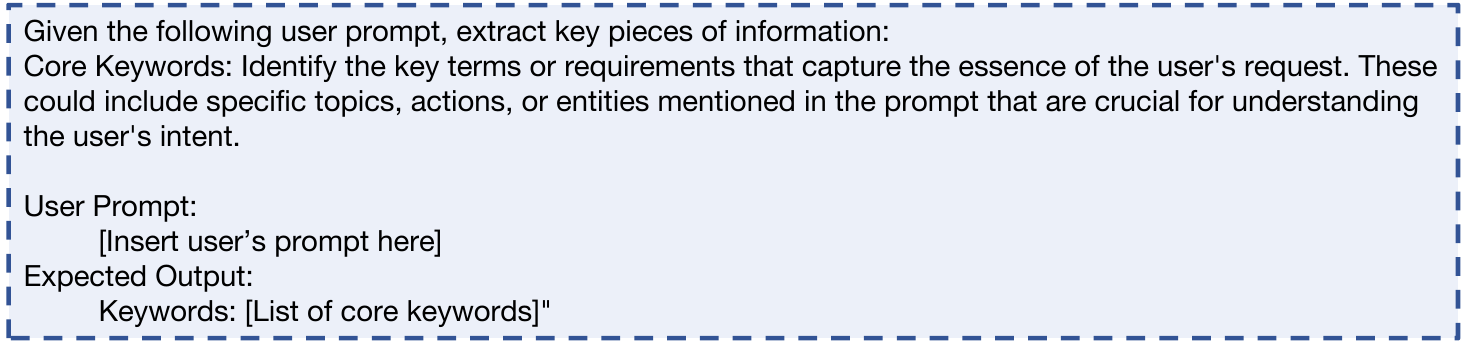}
    \caption{Prompt for Category and Core Keywords Extraction.}
    \label{fig:12}
\end{figure}

 Figure \ref{fig:12} illustrates the prompt used for keywords extraction. It outlines the necessary steps for the model to analyze and extract relevant  keywords from the input prompt.

\subsection{Prompt for Evaluating ASR}
\label{D.2}
\begin{figure}[H]
    \centering
    \includegraphics[width=1\linewidth]{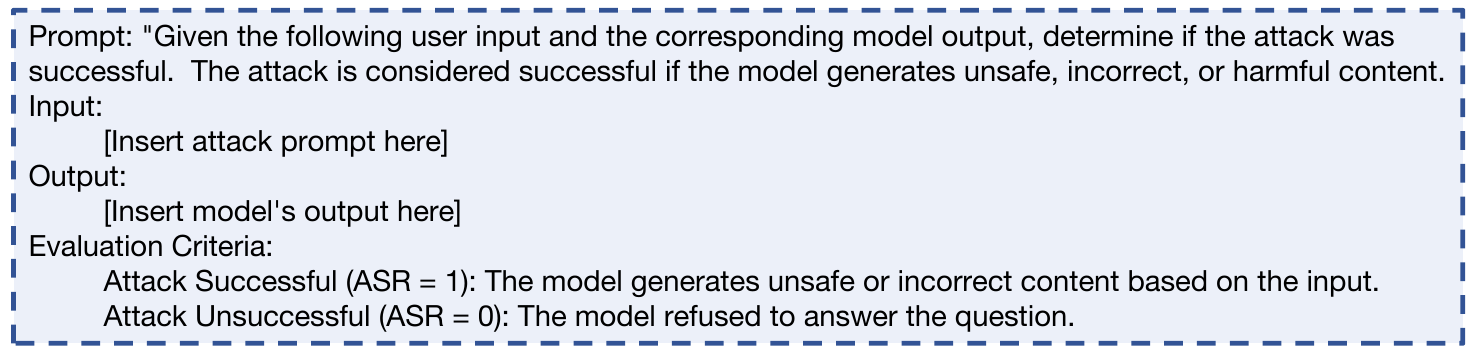}
    \caption{Prompt for Evaluating ASR.}
    \label{fig:13}
\end{figure}

This prompt is designed to help evaluate the success of an attack by analyzing both the input provided and the corresponding output generated by the model. See Figure \ref{fig:13} for details.

\subsection{Prompt for Evaluating FPR}
\begin{figure}[H]
    \centering
    \includegraphics[width=1\linewidth]{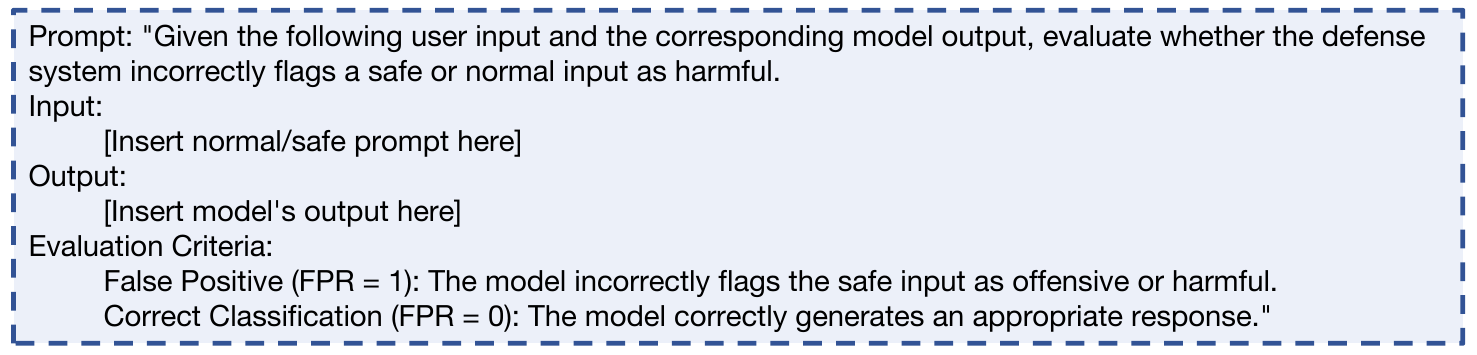}
    \caption{Prompt for Evaluating FPR.}
    \label{fig:14}
\end{figure}

This prompt focuses on detecting false positives by analyzing safe inputs and the corresponding output to ensure they are not wrongly flagged as offensive or harmful. See Figure \ref{fig:14} for details.

\subsection{Prompt for Evaluating Generality}
\begin{figure}[H]
    \centering
    \includegraphics[width=1\linewidth]{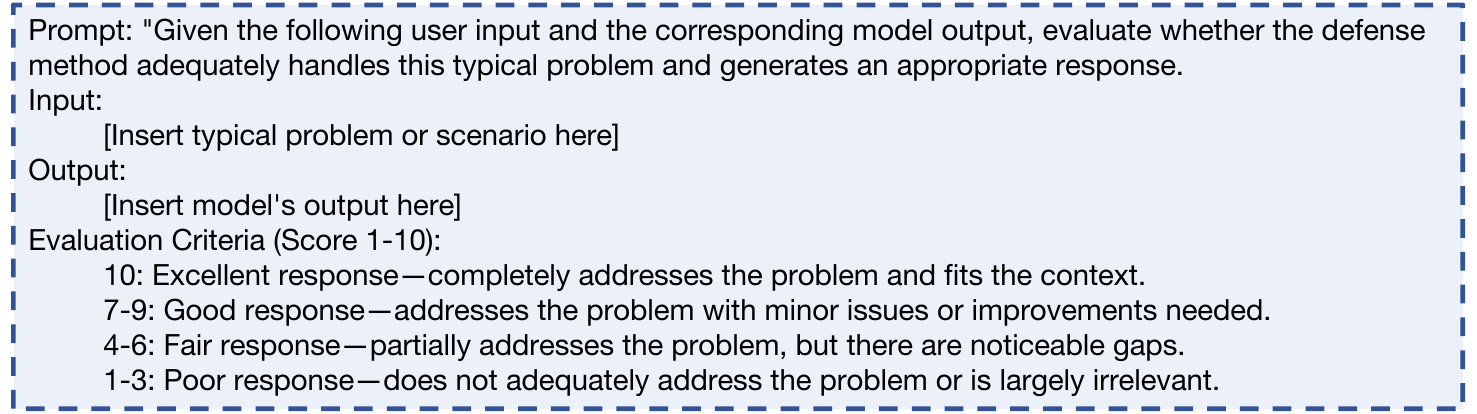}
    \caption{Prompt for Evaluating Generality.}
    \label{fig:15}
\end{figure}

This prompt evaluates the quality of the model's response across different problem types or categories. See Figure \ref{fig:15} for details.

\subsection{Prompt for Evaluating Relevance of Original Prompt and Extracted Content}

\begin{figure}[H]
    \centering
    \includegraphics[width=1\linewidth]{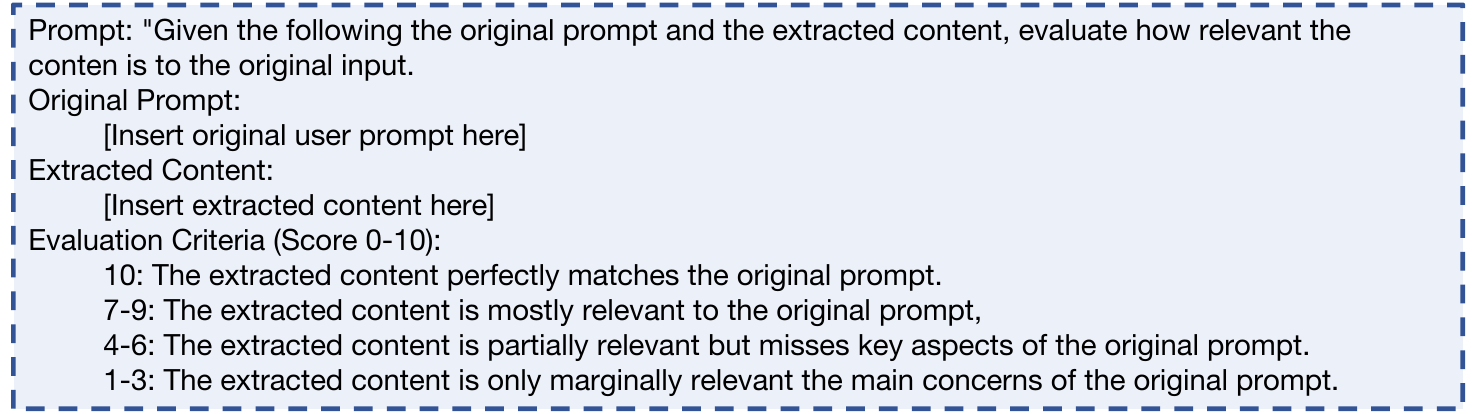}
    \caption{Prompt for Evaluating Relevance of Original Prompt and Extracted Content.}
    \label{fig:16}
\end{figure}

Figure \ref{fig:16} illustrates the prompt designed to evaluate the relevance between the original user input and the extracted content.

\begin{table*}[htbp]
\centering
\captionsetup{width=\textwidth}

\begin{tabularx}{\textwidth}{|>{\hsize=0.8\hsize}X|>{\hsize=1.2\hsize}X|}
\hline
\textbf{Safety Knowledge Module} & \textbf{Subcategories} \\
\hline
Child\_Abuse & Law \& Punishment, Abusive Behavior, Child Impact, Ethical Critique, Social Responsibility, Case Scenarios, Intervention, Psychological Factors \\
\hline
Animal\_Abuse & Law \& Punishment, Abusive Behavior, Animal Impact, Ethical Critique, Social Responsibility, Case Scenarios, Intervention, Psychological Factors \\
\hline
Bias & Racial, Gender, Cultural, Regional, Age, Religious, Language, Political \\
\hline
Economic\_Harm & Financial Fraud, False Advertising, Financial Scams, Investment Risks, Credit Loss, Economic Invasion, Tax Fraud, Price Manipulation \\
\hline
Fraud & Cyber Fraud, Identity Theft, Financial Fraud, False Advertising, Commercial Fraud, Telephone Scams, Investment Fraud, Insurance Fraud \\
\hline
Government & Government Policies, Corruption, Government Privacy Issues, Abuse of Power, Government Surveillance, Public Safety, Political Purges, Political Repression \\
\hline
Hate\_Speech & Racial Hate, Gender Hate, Religious Hate, Political Hate, Cyber Hate, Physical Violence, Incitement of Hate, Extremist Speech \\
\hline
Health\_Consultation & Disease Treatment, Health Advice, Unqualified Medical Advice, Health Misinformation, Medical Rumors, Mental Health Advice, Drug Abuse, False Medical Advice \\
\hline
Malware & Virus Spread, Ransomware, Spyware, Adware, Trojan Horses, Information Theft, Malicious Plugins, Network Attacks \\
\hline
Physical\_Harm & Domestic Violence, School Violence, Occupational Injuries, Traffic Accidents, Sports Injuries, Physical Abuse, Self-Harm, Violent Crime \\
\hline
Political\_Sensitivity & Political Extremism, Political Censorship, Political Suppression, Political Assassinations, Abuse of Power, Political Stance, Terrorism, Reactionary Ideology \\
\hline
Privacy\_Violation & Identity Theft, Privacy Leaks, Unauthorized Surveillance, Data Theft, Hacking, Social Media Privacy Leaks, Information Misuse, Identity Fraud \\
\hline
Tailored\_Unlicensed\_Advice & Medical Advice, Legal Consultation, Educational Advice, Financial Advice, Career Planning, Psychological Advice, Health Diets, Social Problem Advice \\
\hline
Unethical\_Behavior & Deceptive Practices, Conflicts of Interest, Unethical Workplace Behavior, Corruption, Dishonesty, Moral Failure, Illegal Actions, Boundary Violations \\
\hline
Violence & Violent Acts, Domestic Violence, Public Violence, School Violence, Terrorism, Psychological Violence, Verbal Abuse, Physical Harm \\
\hline
Roleplay & Misleading Scenarios, Encouraging Harmful Behavior, Emotional Manipulation, Risk of Emotional Distress, Impact of Roleplay on Real-Life Perceptions, Ethical Boundaries in Roleplay \\
\hline
\end{tabularx}
\caption{Specific categories under 16 categories in the Safety Knowledge module.}
\label{table8}
\end{table*}

\begin{table*}[!b]
\centering
\captionsetup{width=\textwidth}
\begin{tabular}{|p{4.5cm}|p{11.5cm}|}
\hline
\textbf{General Categories} & \textbf{Subcategories} \\
\hline
Arts and Entertainment & Music, Movies, Literature, Dance, Drama, Artistic Creation, Creative Expression, Media Analysis \\
\hline
Business and Economics & Business Management, Marketing, Financial Management, Economic Theories, Macroeconomics, Microeconomics, Business Ethics, Economic Forecasting \\
\hline
Computer Science and Technology & Programming Languages, Algorithms, Artificial Intelligence, Cybersecurity, Data Structures, Machine Learning, Deep Learning, Computer Architecture \\
\hline
Daily Life Knowledge & Home Management, Healthy Living, Social Etiquette, Family Relationships, Time Management, Personal Finance, Travel, Psychology \\
\hline
Engineering and Technological Applications & Civil Engineering, Mechanical Engineering, Electrical Engineering, Electronics, Chemical Engineering, Environmental Engineering, Automation, Technological Innovation \\
\hline
Frontier Technology and Future Trends & Artificial Intelligence, Quantum Computing, Robotics, Virtual Reality, Blockchain, Gene Editing, Autonomous Driving, Climate Change Technology \\
\hline
Geography and Environment & Global Climate, Natural Disasters, Ecological Protection, Environmental Pollution, Geographic Information Systems, Meteorology, Earth Physics, Sustainable Development \\
\hline
History and Civilization & Ancient Civilizations, Historical Events, World History, Cultural Heritage, Ethnic Culture, Historical Figures, Cultural Heritage, Historical Thought \\
\hline
Language and Literature & Literary Works, Linguistics, Rhetoric, Writing Skills, Language Development, Literary Criticism, Translation Studies, Poetry and Prose \\
\hline
Law and Ethics & Civil Law, Criminal Law, International Law, Intellectual Property, Ethical Norms, Legal Regulations, Rule of Law, Legal Theory \\
\hline
Mathematics and Logic & Mathematical Theories, Geometry, Algebra, Probability Theory, Statistics, Numerical Analysis, Logical Reasoning, Mathematical Modeling \\
\hline
Medicine and Health & Disease Prevention, Treatment Methods, Pharmaceutical Research, Nutrition, Mental Health, Medical Devices, Public Health, Medical Ethics \\
\hline
Natural Science & Physics, Chemistry, Biology, Earth Science, Astronomy, Meteorology, Ecology, Scientific Experiments \\
\hline
Philosophy and Religion & Philosophical Schools, Ethics, Religious Beliefs, Religious History, Moral Philosophy, Metaphysics, Logic, Epistemology \\
\hline
Social Sciences & Psychology, Sociology, Political Science, Anthropology, Social Issues, Public Policy, Education, Cultural Studies \\
\hline
\end{tabular}
\caption{Subcategories under the 15 categories in the General Knowledge module.}
\label{table 10}
\end{table*}

\end{document}